# Plasmonic switching in Au functionalized GaN nanowires in the realm of surface plasmon polatriton propagation : A single nanowire switching device

Sandip Dhara,[1,*] Chien-Yao Lu,[2] Kuei-Hsien Chen[2,3,*]

[1] Surface and Nanoscience Division, Indira Gandhi Centre for Atomic Research, Kalpakkam-603102, India.

[2] Institute of Atomic and Molecular Sciences, Academia Sinica, Taipei 106, Taiwan

[3] Center for Condensed Matter Science, National Taiwan University, Taipei 106, Taiwan

*Abstract*

Photoresponse of Au nanoparticle functionalized semiconducting GaN (Au-GaN) nanowires is reported for an optical switching using 532 excitation. Wide band gap GaN nanowires are grown by catalyst assisted chemical vapour deposition technique and functionalized with Au in the chemical route. Au-GaN nanowires show surface plasmon resonance (SPR) mode of Au nanoclusters around 550 nm along with characteristic band for GaN around 365 nm. An optical switching is observed for Au-GaN nanowires with a sub-band gap excitation of 532 nm suggesting possible role of surface plasmon polariton assisted transport of electron in the system. Role of band conduction is ruled out in the absence of optical switching using 325 nm excitation which is higher in energy that the reported band gap of GaN ~ 3.4 eV (365 nm) at room temperature. A finite amount of interband contribution of Au plays an important role along with the inter-particle separation. The switching device is also successfully tested for a single GaN nanowire functionalized with Au nanoclusters. A resistivity value of 0.05 Ω-cm is measured for surface plasmon polariton assisted electrical transport of carrier in the single GaN nanowire.

**Introduction :**

In the last one decade, localized optical field induced extraordinary nonlinear-optical phenomena engrossed tremendous attention in the research community [1-3]. Surface plasmon polaritons (SPP), generated from the interaction of collective oscillation of conduction electrons in noble metal nanoclusters with the excitation of visible light, are reported to propagate near the metal-dielectric interface [4,5]. Coupling of plasmon in noble metal nanoparticles arrays lead to the formation of nano-localized hot spots where incident light intensity can be magnified few orders of magnitude leading to prospective application in sub-micron lithography or imaging at the nanoscale through the use of soft x-rays [6]. As a matter of fact, second harmonic generation (SHG) emanating from surface plasmon resonance (SPR) induced localized field using Au nanoparticles are also vastly accounted in the literature [7,8]. One of the most well known applications is the giant surface-enhanced or noble metal coated tip enhanced Raman scattering (SERS or TERS) enabling single molecule detection [9,10]. Perspective of the interaction of light with matter at the nanoscale is one of the most important elementary issues in optoelectronic applications, e.g., switching devises using all-optical signal processing [3], bio-sensing in the sub-wavelength diffraction limit [11], and two photon absorption (TPA) enhanced SHG [12]. As a matter of fact, role of SPP in the carrier transfer mechanism and TPA enhanced SHG activated band transport of carrier in the conduction mechanism is frequently argued in the literature [13].

We address the issue of carrier transport in reporting an optical switching of Au nanoparticles decorated semiconducting GaN nanowires using sub-band gap excitation assisted with plasmonic coupling of excited photon invoking a possible propagative SPP process in the observed system. Excitation with energy higher than the band gap of the semiconducting GaN nanowire backbone shows absence of optical switching, ruling out band conduction playing any role in the switching device. A single nanowire plasmonic switching device is also demonstrated.

**Experimental Details :**

GaN nanowires are grown using Au catalyst assisted vapour-liquid-solid (VLS) growth process in the chemical vapour deposition technique at 900 °C with $NH_3$ (10 sccm) as reactant gas. Nanaowires were hydroxylated in acidic solution of $H_2SO_4$ (0.12 M) and $HNO_3$ (0.52 M) at room temperature (RT) for 1 h for introducing hydroxyl terminated groups on nanowire surface. Then surfaces of the hydroxilated nanowires were chemically functionalized by organosilane MPTS (3-mercaptopropyl trimethoxysilane) for attaching Au nanoparticles on the thiol terminated end. The sample was subjected to multiple cleaning and rinsing to make sure that no physisorbed Au nanoparticle remained on the nanowire surface. The growth of these nanowires and functionalization process is detailed in the earlier report [14].

Photoconductivity studies were carried out using diode lasers with emission length of 532 (252 mW) and 325 nm (40 mW) using a probe station equipped with an optical microscope and two manipulators at room temperature. Under a magnified image in the microscope, each manipulator connected with a tungsten needle (tip size ≈2 μm) could be precisely controlled in their probing position on the micrometer-sized nickel electrodes of the device. A semiconductor characterization system (Keithley model 4200-SCS) was utilized to source the dc bias and to measure the current.

**Results and Discussion :**

The Au nanoparticle decorated GaN (Au-GaN) nanowires were physically transferred by sonication in ethanol $C_2H_5OH$ and then by drop casting method and subsequent infrared heating to patterned substrate with Au as metal contact pads and $SiN_x$ as an insulating layer. Thermal conductivity of these substrates are reasonably high not to affect the electrical measurements under optical illumination with laser. It is shown in the scanning electron microscopic (SEM; FEI NOVA-600) image for ensembled brunch of nanowires (supplementary information Fig. S1(a)) and single nanowire (supplementary information Fig.S1(b)) before making electrical contact pads. Backscattered electron images (BEI) using trium-aluminum garnet (YAG) scintillator detector in the field emission scanning electron microscope (FESEM, JEOL JSM-6700F) of brunch of nanowires (Fig. 1(a)) and single nanowire (Fig. 1(b)) using FESEM show attached Au nanoparticles on GaN nanowires. A typical Au-GaN nanowire was examined using transmission electron microscopy (TEM; by dispersing the nanowires on lacey carbon covered Cu grid, using JEOL, JEM-4000EX) in the bright field image with typical image showing ~ 5 nm Au nanoparticles on ~40 nm GaN nanowire (Fig. 1(c)). UV-Visible (Jasco V-650) reflectance studies in these Au-GaN nanowire assembly (Fig. 1(d)) on Si substrate show change in slopes, indicating absorption in material at around 340-370 nm (3.35-3.65 eV) and 510-580 nm (2.1-2.4 eV) corresponding to the band edge of these GaN nanowires ~ 365 nm (3.4 eV) reported elsewhere [14], and SPR absorption peak of Au around ~550 nm (2.25 eV) [15], respectively. A dual beam focused ion beam (FIB; FEI NOVA-600) system with a SEM and a 30 KeV $Ga^+$ FIB at a fluence of 1E10 ions·$cm^{-2}$ were used for metallization in Au-GaN nanowires with different configurations of brunch of nanowires (Fig. 2(a); inset showing close look to the bunch) and a single nanowire of diameter ~ 50 nm and length ~1 μm (Fig. 2(b)) by adopting ion beam assisted disintegration of trimethyl cyclopentadienyl platinum ($C_9H_{16}Pt$) for photoconductivity studies.

Photoresponse measurement based on the Au-GaN nanowire system was conducted alternatively under light illumination of different wavelengths and dark condition. Prior to the experiment, background resistance based on blank substrate ($SiN_x$-coated Si(100)) under light illumination (532 nm and 325 nm) was also investigated as a reference (not shown in figure). The background resistance from the blank substrate showed very high resistance and no photoresponse was observed upon exposure of laser light illumination. The resistance of the Au-GaN nanowires in ensemble (Fig. 3(a)) exhibits nearly 90% change in the resistance with the exposure of sub-band gap excitation wavelength of 532 nm. We also observe an appreciable ~ 80% change in the resistance for a single Au-GaN nanowire (Fig. 3(b)) with the exposure of sub-band gap excitation wavelength of 532 nm. Lower resistance in the single nanowire than that for the ensemble of nanowires may be because of possible random connectivity among the nanowires in the ensemble. On the other hand, a noticeable noise was observed in the single nanowire photoresponse measurements. It may be correlated to contact resistances in the single nanowire device. In this regard, we also like to state that instead of resistivity we have measured pure resistance in these systems of single and ensemble of nanowires for comparison purpose, as the resistivity measurement will be completely erroneous for randomly connected ensembled nanowires. The resistivity of the single nanowire with length ~ 1 μm, diameter ~ 50 nm (Fig. 2(b)), however can be measured as ~ 0.3 Ω-cm in the dark and 0.05 Ω-cm in the 532 nm illumination conditions. The dark resistivity is comparable to unintentionally doped GaN film and NW systems [16,17]. However, absence of switching activity with 325 nm excitation (not shown in figure) rules out band related conduction mechanism as band gap of the semiconducting GaN is reported at lower energy ~ 3.4 eV (365 nm) at room temperature. Thus, a TPA assisted SHG generation in the localized field due to coupling of SPR of Au nanostructures [18], and its role in band conduction is also negated simultaneously. The optical switching of the Au-GaN nanowire system may primarily due to propagative SPP involving electrical carriers originating from the strong coupling of SPR of Au nanoparticles in the form of conduction electron of noble metal Au. It may be relevant here to mention that the SPP assisted photoresponse does not occur across the band. Instead the photocurrent is due to the transport of conduction electrons belonging to Au as carrier for the electrical conduction process. Thus the reduced resistivity value of 0.05 Ω-cm for single GaN nanowire in the illumination of 532 nm may be the characteristic to SPP propagated conduction system where SPR play the role of supplying the conduction electrons of Au as carriers for the photoresponse. We have evidenced SPR related absorption of the Au-GaN nanowire system (Fig. 2(d)). As discussed earlier, the SPR coupled electromagnetic radiation, however can also travel several tens of nanometers making resonating electron travel across these nanowire system [19,20]. Finite value of interband contribution in Au may play a definitive role in the SPP assisted process [15]. Au nanoparticles in these nanowires are also distributed in sufficient close proximity (Fig. 2) to take part in the conduction process [21]. Thus, the conduction is because of SPP assisted electron transport supported by propagating resonating electromagnetic radiation.

**Conclusion :**

In conclusion, an optical switching for Au nanoparticle decorated semiconducting GaN nanowires is demonstrated for sub-band gap (532 nm) excitation with surface plasmon resonance peak of Au around 550 nm. A single nanowire plasmonic switching device is also demonstrated using the same 532 nm excitation. Conduction process is conceived as plasmon polariton assisted transport of electron with propagative resonating electromagnetic radiation coupled to Au nanoclusters generating the carriers which are responsible for the observed photoresponse. Resistivity measurements in the single nanowire shows lowering of resistivity from the bulk value of 0.3 $\Omega$-cm (measured at dark) to 0.05 $\Omega$-cm for the 532 nm illumination as a measure of surface plasmon polariton assisted electrical conduction process. Role of two photon assisted second harmonic generation and its role in band conduction for the switching action is ruled out in the absence of excitation with higher energy than that of the band gap of the semiconductor backbone.

**Acknowledgements**

We thank C. P. Chen, A. Ganguly, C. W. Hsu, and C. T. Wu of CCMS, Taiwan for their contribution in performing the experiments. We are greatly in debt to L. C. Chen of CCMS, Taiwan for useful discussion.

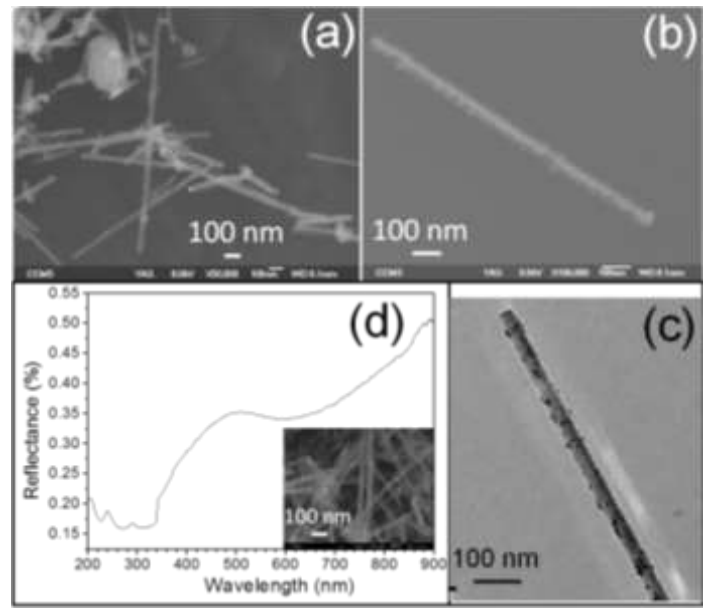

**Fig. 1.** Backscattered electron images of (a) ensemble and (b) a single nanowire showing Au nanoparticles on the GaN nanowire system. (c) typical transmission electron microscopic image of a single GaN nanowire decorated with Au nanoparticles and (d) reflectance spectra of Au functionalized GaN nanowires on Si showing absorption in the range of 340-370 nm and 510-580 nm. Inset shows the backscattered electron images of the sample.

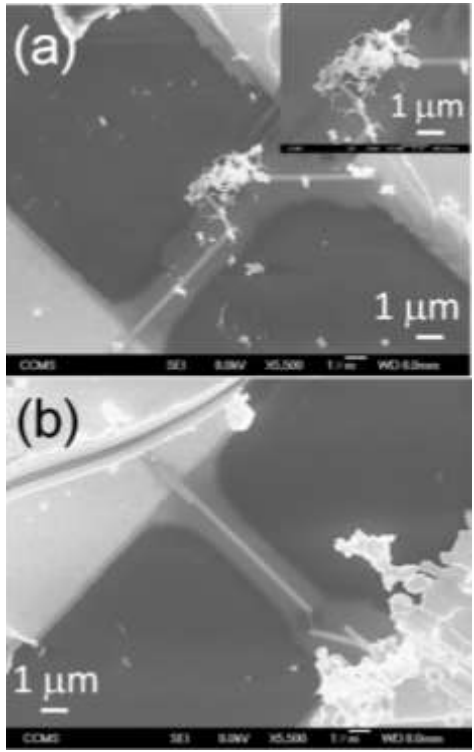

**Fig. 2.** Focused ion beam assisted Pt contact electrodes in (a) ensemble (b) single GaN nanowire system

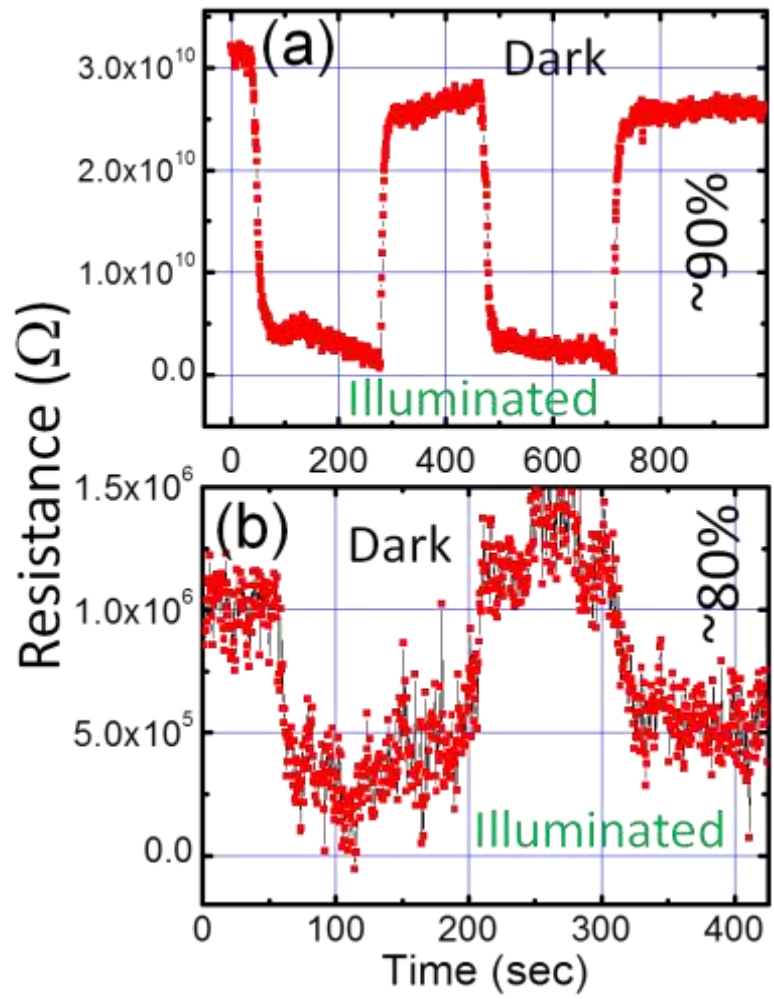

**Fig. 3.** Photoresponse in a) ensemble b) single GaN nanowire samples with periodical dark and 532 nm illumination conditions.

**Supporting Information**

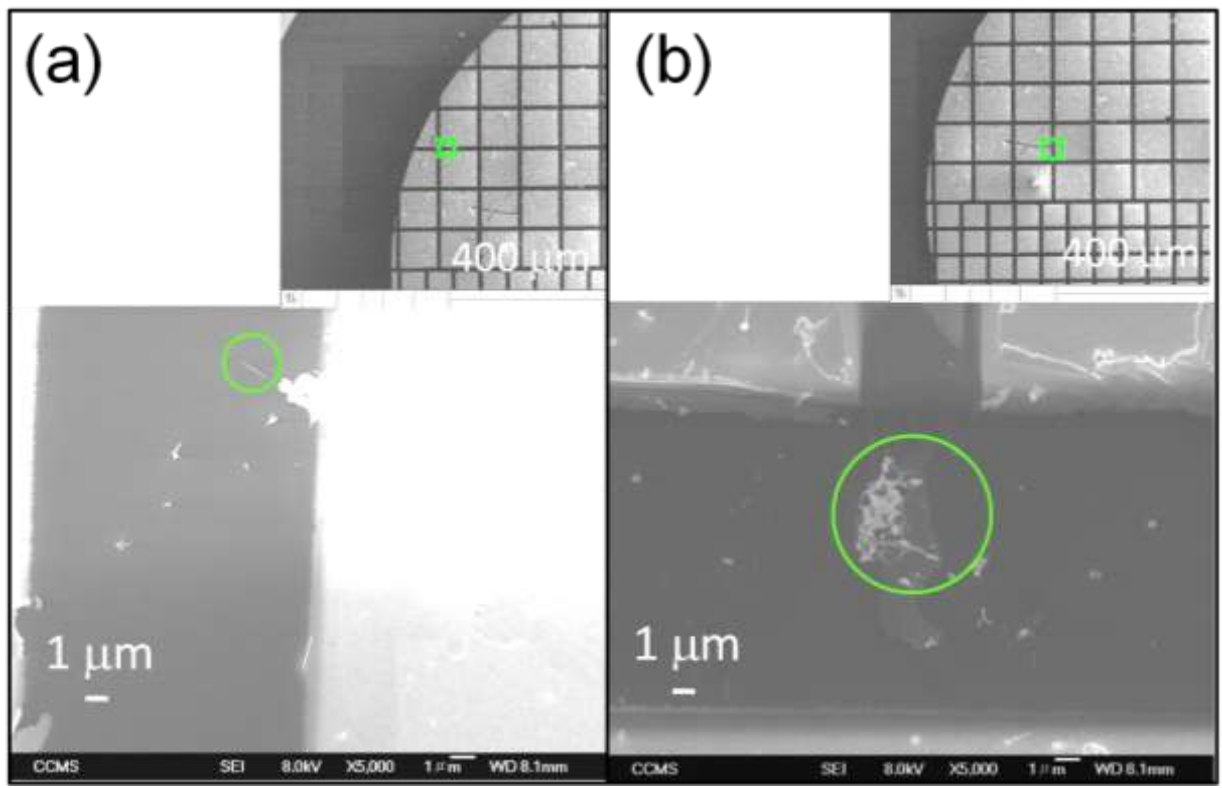

Fig. S1. Scanning electron microscopic images of the (a) single and (b) ensemble of GaN nanowire along with Au contact pad at different magnifications.